\documentclass[aps,10pt,american,prl,twocolumn,groupedaddress, superscriptaddress]{revtex4-1}
\usepackage[latin9]{inputenc}
\usepackage{color}
\usepackage{placeins}
\usepackage{amsmath}
\usepackage{amssymb}
\usepackage{cancel}
\usepackage{graphicx}
\usepackage{wasysym}
\usepackage{multirow}
\usepackage{hyperref}
\usepackage{comment}

\graphicspath{{figs/}}

\makeatletter

\newcommand{\noun}[1]{\textsc{#1}}

\@ifundefined{textcolor}{}
{
 \definecolor{BLACK}{gray}{0}
 \definecolor{WHITE}{gray}{1}
 \definecolor{RED}{rgb}{1,0,0}
 \definecolor{GREEN}{rgb}{0,1,0}
 \definecolor{BLUE}{rgb}{0,0,1}
 \definecolor{CYAN}{cmyk}{1,0,0,0}
 \definecolor{MAGENTA}{cmyk}{0,1,0,0}
 \definecolor{YELLOW}{cmyk}{0,0,1,0}
 \definecolor{armygreen}{rgb}{0.55, 0.73, 0.0}
}

\newcommand{\nuNull}{2.30 \textendash 2.36}

\makeatother

\begin{document}

\title{Criticality of two-dimensional disordered Dirac fermions in the unitary class and universality of the integer quantum Hall transition}

\date{\today}

\author{Bj\"orn Sbierski}
\affiliation{Department of Physics, University of California, Berkeley, California 94720, USA}

\author{Elizabeth J. Dresselhaus}
\affiliation{Department of Physics, University of California, Berkeley, California 94720, USA}

\author{Joel E. Moore}
\affiliation{Department of Physics, University of California, Berkeley, California 94720, USA}
\affiliation{Materials Sciences Division, Lawrence Berkeley National Laboratory, Berkeley, California 94720, USA}

\author{Ilya A. Gruzberg}
\affiliation{Ohio State University, Department of Physics, 191 West Woodruff Ave, Columbus OH, 43210}

\begin{abstract}
Two-dimensional (2D) Dirac fermions are a central paradigm of modern condensed matter physics, describing low-energy excitations in graphene, in certain classes of superconductors, and on surfaces of 3D topological insulators. At zero energy $E=0$, Dirac fermions with mass $m$ are band insulators, with the Chern number jumping by unity at $m=0$. This observation lead Ludwig \emph{et al.}~[Phys. Rev. B {\bf 50}, 7526 (1994)]
to conjecture that the transition in 2D disordered Dirac fermions (DDF) and the integer quantum Hall transition (IQHT) are controlled by the same fixed point and possess the same universal critical properties.
Given the far-reaching implications for the emerging field of the quantum anomalous Hall effect, modern condensed matter physics and our general understanding of disordered critical points, it is surprising that this
conjecture has never been tested numerically. Here, we report the results of extensive numerics on the phase diagram and criticality
of 2D-DDF in the unitary class. We find a critical line at $m=0$, with energy-dependent localization length exponent. At large energies, our results for the DDF are consistent with state-of-the-art numerical results $\nu_\text{IQH} = 2.56 \textendash 2.62$ from models of the IQHT. At $E=0$ however, we obtain $\nu_0 =\nuNull$ \textit{incompatible} with $\nu_\text{IQH}$. This result challenges conjectured relations between different models of the IQHT, and several interpretations are discussed.
\end{abstract}

\maketitle


{\it Introduction.} The integer quantum Hall effect appears when a two-dimensional (2D) electron gas is placed in a strong perpendicular magnetic field. Without disorder, the electron eigenstates form Landau levels, and each filled level contributes unity to  the total Chern number $C$. Disorder is essential for experimental observation of the (dimensionless) quantized Hall conductivity $\sigma_{xy} = C$; it broadens the Landau levels into bands and localizes eigenstates on a scale $\xi(E)$ that diverges as a power law at a critical energy $E_{c}$ \citep{Huckestein1995}, $\xi(E)\sim |E-E_{c}|^{-\nu_\text{IQH}}$.
For Fermi energies $E\! \neq \!E_c$ and system sizes $L\!\gg\!\xi(E)$ the Hall conductivity is quantized. The integer quantum Hall transition (IQHT) at $E = E_c$ is the most studied Anderson transition \citep{EversMirlin:review} because of its conceptual simplicity, low dimensionality, and experimental relevance. However, critical properties at the IQHT are notoriously difficult to compute analytically; they are mostly known from numerical studies which employed the Chalker-Coddington (CC) network model \citep{Chalker-Percolation-1988, Kramer-Random-2005, Obuse-Boundary-2008, Evers-Multifractality-2008, Slevin-Critical-2009, Obuse-Conformal-2010, Amado-Numerical-2011, Fulga2011a, Obuse-Finite-2012, Slevin-Finite-2012, Nuding-Localization-2015}, microscopic continuous \cite{Ippoliti-Integer-2018, Zhu-Localization-length-2019}, lattice \cite{Fulga2011a, Ippoliti-Integer-2018, Zhu-Localization-length-2019, Puschmann-Integer-2019, Puschmann-Edge-2020-arXiv}, and Floquet Hamiltonians \cite{Dahlhaus-Quantum-2011}. In recent works, the critical properties agree between models, indicating universality of the IQHT. They include the localization length exponent $\nu_\text{IQH}=2.56 \textendash 2.62$ and the leading irrelevant exponent  $y\simeq0.4$ (with large error bars). At criticality, $y$ describes the approach of the dimensionless quasi-1D Lyapunov exponent $\Gamma$  to its limiting value at infinite system size $\Gamma_{0}^\text{IQH}=0.77 \textendash 0.82$ \citep{Obuse-Boundary-2008, Evers-Multifractality-2008, Slevin-Critical-2009, Amado-Numerical-2011, Obuse-Finite-2012, Slevin-Finite-2012, Nuding-Localization-2015, Puschmann-Integer-2019}. A similar exponent $y$ was found for the average conductance $\overline{g}$ of a square sample with limiting value $\overline{g}_\text{IQH} = 0.58 \textendash 0.62$ \citep{Wang-Critical-1996, Schweitzer2005}. For ongoing analytical work on the IQHT, see \citep{Bondesan-Pure-2014, Bondesan-Gaussian-2017, Zirnbauer2019} and the discussion below. The IQHT has also been discussed recently in relation to exotic topological superconductor surface states \citep{SbierskiPRX2020}.

\begin{figure}
\centering
\includegraphics{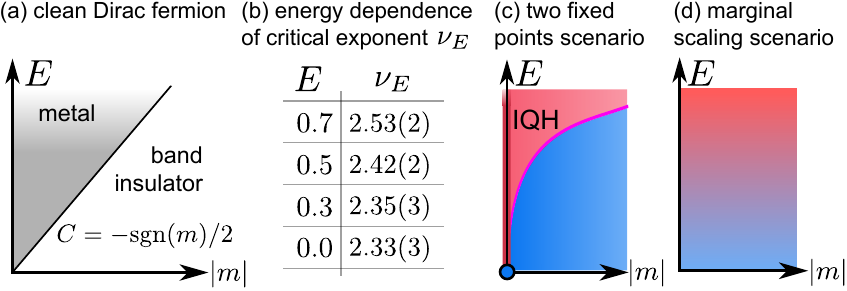}
\caption{{\small{}Schematic phase diagram for 2D Dirac fermions. (a) Clean case: A metal intervenes between two band insulators with different Chern numbers $C$ at $|m|>|E|$. With disorder in the unitary class, the metal localizes except on the critical line $m=0$ separating topologically distinct Anderson insulators. (b) The critical exponent $\nu_E$ is found to vary significantly with energy. The two fixed points scenario (c) explains this as a result of a crossover, while the marginal scaling scenario (d) would be compatible with a smooth evolution of effective critical exponents.
}}
\label{fig:phaseDiagram}
\end{figure}

A longstanding conjecture by Ludwig \emph{et al.}~\citep{Ludwig1994} states that the IQHT fixed point also controls the criticality of 2D disordered Dirac fermions (DDF). The clean Dirac Hamiltonian is
\begin{equation}
H_{0} = \hbar v \left(-i\sigma_{x} \partial_{x} - i\sigma_{y} \partial_{y} \right) + m \sigma_{z},
\label{eq:H0}
\end{equation}
with Pauli matrices $\sigma_\mu$, mass $m$, and velocity $v$. The spectrum of $H_0$ has a gap $2|m|$ symmetric around $E=0$. For Fermi energies $E$ within the gap, the system is a band insulator with half-integer quantized $\sigma_{xy} = C(m) = -\frac{1}{2}\mathrm{sgn}(m)$ \citep{Ludwig1994}, see Fig. \ref{fig:phaseDiagram}(a). If the Dirac fermion is regularized on a lattice as in the Haldane model \citep{Haldane1988} or Eq.~\eqref{eq:H0_Lattice} below, $H_0$ only describes the low-energy excitations near a certain point in the Brillouin zone. Bloch states elsewhere contribute another 1/2 to $C$,
such that $|\sigma_{xy}|$ jumps between zero and one as $m$ changes sign.

With $m$ taking the role of energy, the superficial similarity of this transition to the IQHT motivated Ludwig \emph{et al.~} \citep{Ludwig1994} to consider the effects of disorder in the unitary symmetry class \cite{Zirnbauer-Riemannian-1996, Altland1997},
\begin{equation}
H = H_{0} + \sum_{\mu=0,x,y,z} U_{\mu}(x,y) \sigma_{\mu}.
\label{eq:H}
\end{equation}
The random scalar ($U_{0}$) and vector ($U_{x,y}$) potentials, and the random part of the mass ($U_{z}$) are taken to be independent Gaussian fields with the correlators $\overline{U_{\mu}(\mathbf{r}) U_{\nu}(\mathbf{r}^{\prime})} \!=\! \delta_{\mu\nu} K_\mu(|\mathbf{r}^{\prime}-\mathbf{r}|)$ and zero mean. Time reversal changes the sign of $m+U_z$, connecting two equally likely members of the statistical ensembles with opposite values of $m$, and the transition in the disordered model happens at $m=0$. Due to the absence of an extended 2D metal phase in the unitary class, all eigenstates of $H$ with $|m| > 0$ are expected to be localized with the localization length $\xi(m) \sim |m|^{-\nu_E}$, with a possibly $E$-dependent critical exponent $\nu_E$.

Although model \eqref{eq:H} is not solvable analytically, the conjecture \cite{Ludwig1994} $\nu_\mathrm{E=0}=\nu_\text{IQH}$ was based on a semiclassical argument that leads to the CC-model. Another argument \cite{Ho1996} considers the clean CC-model and finds a Dirac spectrum, but the inclusion of disorder is  uncontrolled. In the supplemental material (SM) \footnote{See Supplemental Material [url] for non-rigorous arguments for the equivalence of IQHT and DDF criticality, details on the fitting procedure for quasi-1d Lyapunov exponents, on the alternative scaling observable, Landauer conductance and Drude conductivity. The supplemental material contains references \citep{Klumper2019,Janssen-1999,Klesse-2001,Gruzberg-Classification-2013,Jovanovic1998}}, we review these arguments and identify their possible flaws.

Despite the importance of the 2D Dirac model in modern physics, the conjectured emergence of IQHT criticality in DDF was never checked numerically. Here, we address this issue with extensive simulations employing different microscopic models and scaling observables. We start with the continuum model \eqref{eq:H}
and use the transfer matrix (TM) approach in quasi-1D (q1D) geometry to find the critical behavior near the line $m=0$ in the $m$-$E$ plane, see Fig. \ref{fig:phaseDiagram}(b). At large $E$ our results are consistent with $\nu_E=\nu_\text{IQH}$, but as $E$ is lowered, the critical exponent decreases towards $\nu_{E=0} = 2.33(3)$ still close to, but strikingly \emph{incompatible} with $\nu_\text{IQH}$.
We corroborate our $E=0$ results in a lattice model of DDF, employing an alternative 2D scaling observable \cite{Fulga2011a}.

In the experimental literature, a quantized non-zero $\sigma_{xy}$ in the absence of an external magnetic field is known as the quantum anomalous Hall effect \cite{Chang-QAH-experiment-2013, Liu-QAH-review-2016, Serlin-QAH-MoireGraphene-2020}. Recent efforts \cite{Chang-QAH-scaling-experiment-2016, Kawamura-QAH-scaling-experiment-2020} have been directed to the critical scaling at the topological phase transition in question, however, the error bars on the resulting exponents are still large.


{\it Continuum model and disorder-induced length scale.} We start with Hamiltonian \eqref{eq:H} at $E=0$ and smooth disorder,
$K_\mu(r) = W^{2} e^{-r^{2}/2a^{2}}/ 2\pi$. We use the disorder correlation length $a$ and $\hbar v/a$ as units of length and energy so that the dimensionless disorder strength $W$, taken to be the same for all four disorder fields, is the bare energy scale in the model. The mean free path $l_W$ equals the quasiparticle decay time, $l_W \equiv -1 / \mathrm{Im} \, \Sigma_{\uparrow \uparrow}(0, 0)$ defined in terms of the disorder-averaged Green function $\overline{G(\mathbf{k},\omega)} = [\omega - H_0(\mathbf{k}) - \Sigma(\mathbf{k},\omega)]^{-1}$. For weak disorder $W \ll 1$, a perturbative renormalization group (RG) \citep{Ludwig1994, Schuessler2009} gives, for $m=0$, $l_W \propto e^{c/W^{2}}$, with $c = O(1)$. To ensure that our system sizes $L \gg l_W$, we work with strong disorder $W \geq 1.5$ where a numerically exact method \citep{Sbierski2019} yields $l_{W=1.5} = 1.54$. We also observe that for $k l_W > 1$, the peaks in the spectral function $A(\mathbf{k},\omega) = - \frac{1}{\pi} \mathrm{tr}\, \mathrm{Im}\, \overline{G(\mathbf{k},\omega)}$ occur at frequencies $\omega \simeq \pm \hbar vk$, i.e. the velocity $v$ is almost un-renormalized. We conclude that for $W=1.5$, system sizes $L \gtrsim O(10)$ are large enough to exhibit disorder-dominated physics.

{\it Lyapunov exponent (LE).} A common method to analyze critical behavior in disordered systems employs the self-averaging LEs $\gamma_i$ in a quasi-1D geometry with length $L_{x} \rightarrow \infty$ \citep{Kramer1993}. The smallest $\gamma_i>0$ (the inverse of the 1D localization length) gives the scaling variable $\Gamma = \gamma L_{y}$, which increases (decreases) with width $L_{y}$ in a localized (extended) phase and is scale-invariant at a critical point. Following Ref.~\citep{Amado2011}, we use finite $L_x=O(10^5)$ and find $\Gamma$ as the average over hundreds of disorder realizations, see SM for details.

The eigenvalue problem for the DDF \eqref{eq:H} can be rewritten as $\partial_{x} \psi(x,k_{y}) = f(\psi(x,k_{y}^{\prime}))$. The right hand side contains scattering between transversal wavevectors $k_y$ but is local in $x$, which allows us to express the TM in exponential form. We impose periodic boundary conditions (BC) in the $y$ direction.
We discretize the $x$ direction and stabilize the TM multiplication by repeated QR-decompositions \citep{Huckestein1995} (to obtain $\Gamma$) or via a scattering matrix \citep{Bardarson2007} (for the conductance of moderately sized systems). Both methods are numerically exact and faithfully treat model \eqref{eq:H} without band bending or node doubling. The only approximations are related to the cutoff $|k_{y}| \leq k_\text{max}$ and the $x$-discretization. The associated length scales (taken equal) were chosen much smaller than $a$, and results are converged with respect to these parameters.

\begin{figure}
\centering
\includegraphics{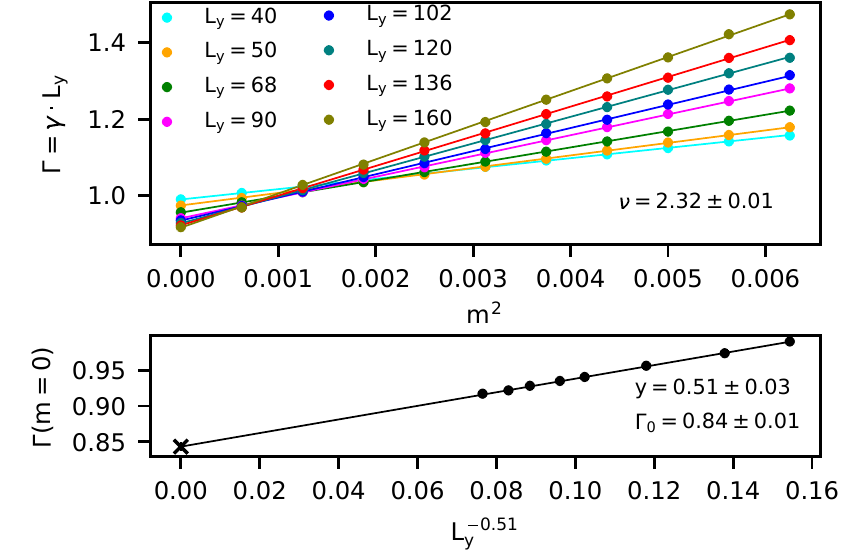}
\caption{Top: LEs $\Gamma$ for $E=0$ and $W=1.5$ as functions of $m^2$. The relative error is $\leq0.2\%$, error bars are smaller than the dots. Solid lines denote the best fit [Eq.~\eqref{eq:simpleScaling}] with fit parameters as given in the panels. Bottom: Closeup at criticality ($m=0$) with extrapolation to infinite system size determining $\Gamma_0$ (cross).}
\label{fig:GammaScaling_W1.5_kF0}
\end{figure}

Results for the dimensionless LE $\Gamma$ at $E=0$, $W=1.5$, various masses $m$ and system widths $L_{y}$ are presented in Fig. \ref{fig:GammaScaling_W1.5_kF0}.
The solid lines are fits to the scaling function
\begin{equation}
\Gamma(m,L_y)=\Gamma_0+\alpha_{01}L_y^{-y}+\alpha_{20}m^2L_y^{2/\nu},
\label{eq:simpleScaling}
\end{equation}
which is the lowest-order polynomial ansatz allowed by symmetry, including an irrelevant contribution. The fit gives the following critical properties: 
\begin{align}
\nu_{E=0} &= 2.32(1), & y &= 0.51(3), & \Gamma_0 &= 0.84(1),
\end{align}
the number in parentheses denotes one standard deviation. In the SM, we give a detailed account for the fitting procedure and show its stability with respect to higher order terms in Eq.~\eqref{eq:simpleScaling} and a removal of data points for large $m$ and small $L_y$. There, we also present data for an increased disorder strength $W=2.0$, which yields $\nu_{E=0} = 2.31(2)$, $y = 0.51(3)$ and $\Gamma_0 = 0.84(1)$ compatible with anticipated disorder-independent critical properties.


{\it Lattice model and alternative scaling observable.} We now confirm the value of $\nu_{E=0}$ using a square-lattice regularization of the DDF allowing access to an alternative scaling observable introduced by Fulga \emph{et al.}~\citep{Fulga2011a}. In momentum space, the clean model reads \citep{Qi2006}
\begin{equation}
H_{0}^L \!= \sigma_{x} \sin k_{x} + \sigma_{y} \sin k_{y} + \sigma_{z}
(m-2 + \cos k_{x} + \cos k_{y}),
\label{eq:H0_Lattice}
\end{equation}
where lattice constant and energy scale have been set to unity. For $|\mathbf{k}| \ll 1$, this model reduces to Eq.~\eqref{eq:H0}, with a topological transition at critical $m=m_c=0$ where $C$ changes by 1, but band bending is important for $k, E \gtrsim 1$. We add on-site disorder potentials, $V = \sum_{\mathbf{r}_{i},\mu} U_{\mu}(\mathbf{r}_{i}) \sigma_{\mu}$ with $U_{\mu}(\mathbf{r}_{i})$ uniformly drawn from the interval $[-w/2,w/2]$ independently for each lattice site $\mathbf{r}_{i}$ and $\mu=0,x,y,z$. Transport calculations use the \noun{kwant} package \citep{Groth2014} and employ two identical leads attached at the left and right boundaries of the system, represented by decoupled 1D chains extending in the $x$-direction:
\begin{equation}
H_{Lead}(k_{x},k_{y})=\sigma_{x}\sin k_{x}+\sigma_{z}\left(1+\cos k_{x}\right).\label{eq:H_lead}
\end{equation}
The lattice model \eqref{eq:H0_Lattice} has no symmetry that ensures $m_c=0$ in the presence of disorder. However, the Dirac node energy is not renormalized away from $E=0$. The reason is that the eigenenergies come in pairs $\pm E$. This symmetry carries over to the disorder averaged density of states as long as the average potential disorder $\overline{U_0} = 0$.

To determine the exponent $\nu_{E=0}$, we consider the reflection matrix $r(\phi)$ of the left lead as a function of the phase $\phi$ of twisted BC in the $y$ direction. For a given disorder realization, the $m_c$ occurs when there exists a $\phi$ such that $r(\phi)$ has a zero eigenvalue and $\det r(\phi)=0$. Fulga \emph{et al.}~\citep{Fulga2011a} showed that a scaling observable $\Lambda$ can be obtained by working with generalized twisted BC $\psi_{x,y=L-1} = z\, \psi_{x,y=0}$ for all $x = 0, 1, ..., L-1$, and $z \in \mathbb{C}$. Now, $\det r(z)$ has zeros $z_{0}$ even for $m\neq m_c$ but with $|z_{0}|\neq1$. For the $z_0$ closest to the unit circle, $\Lambda = \overline{\mathrm{log}|z_{0}|}$ measures the distance to criticality $\Lambda=0$. For the CC-model, scaling of $\Lambda$ with system size $L$ was demonstrated in Ref.~\cite{Fulga2011a}, reporting $\nu=2.56(3)$ compatible with results from the TM method.

\begin{figure}
\centering
\includegraphics{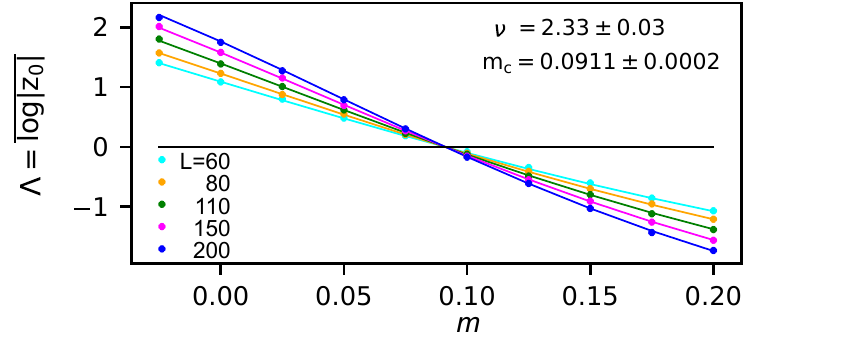}
\caption{{\small{}Scaling plot of the variable $\Lambda$ for the model \eqref{eq:H0_Lattice} at $E=0$ and disorder strength $w=2.5$. Dots represent averages over at least $10^{4}$ disorder realizations, and the solid curves are fits described in the SM.}}
\label{fig:FulgaScaling}
\end{figure}

We computed $\Lambda$ for the lattice DDF $H_0^L+V$ for $m$ around $0$, $w=2.5$ and system sizes between $L=60$ and $200$, see Fig. \ref{fig:FulgaScaling} for the results and the SM for details of the fit. We find $\nu_{E=0}=2.33(3)$ in agreement with the result for the continuum model. Notably, the observable $\Lambda$ shows no discernible corrections to scaling, which allows us to omit the irrelevant terms in the scaling function for $\Lambda$. Repeating the analysis for $w=2.25$ and $2.75$ (not shown) yields compatible $\nu$ within the given error bars.

{\it Results for finite energy ($E>0$).} We now consider the continuum model \eqref{eq:H} with smooth disorder at finite energy $E>0$ ($E<0$ is related by the statistical $E\rightarrow-E$ symmetry). In the SM, we present scaling results for the LE $\Gamma$ for $E=0.3,0.5,0.7$ at disorder strength $W=2$. As in the $E=0$ case we find localizing behavior for any $m \neq 0$. The exponents $\nu_E$, see Fig.~\ref{fig:phaseDiagram}(b), increase monotonically with $E$ towards $\nu_{E=0.7} = 2.53(2)$, significantly different from $\nu_{E=0}$. Other critical properties ($\Gamma_0$ and $y$) do not seem to vary significantly with $E$.

\begin{figure}
\noindent \centering{}\
\includegraphics{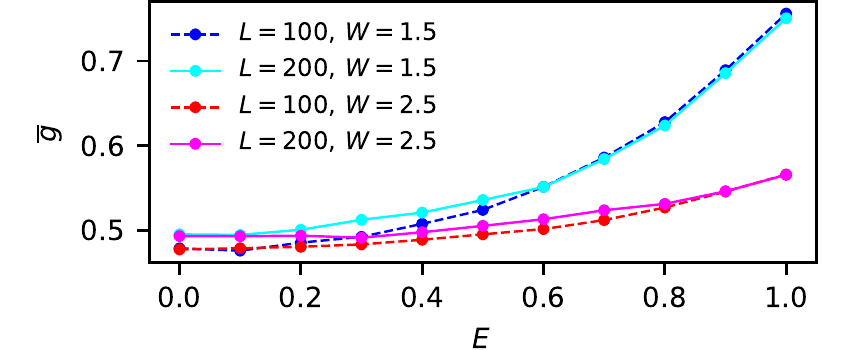}
\caption{{\small Critical Landauer conductance $\overline{g}$ of square samples at $m=0$, disorder strengths $W=1.5$ and $2.5$, size $L=100,200$ and periodic BC in transversal direction averaged over at least $10^4$ disorder realizations.}}
\label{fig:gsq_vs_kF}
\end{figure}

To further probe the critical line $m=0$, we compute the critical Landauer conductance $g$ of $L \times L$ systems with periodic BC in the $y$ direction, and metallic leads modeled as highly doped Dirac nodes \citep{Tworzydlo2006a}. The distribution of $g$ and its moments are expected to be scale-invariant and universal \citep{Kramer-Random-2005, EversMirlin:review}, for $E=0$ it is shown in the SM. In Fig.~\ref{fig:gsq_vs_kF} we present the average conductance $\overline{g}$. We observe that for $E \lesssim 0.3$, $\overline{g} \simeq 0.5$ is almost independent of the disorder strength and $E$, which we interpret as evidence of proximity to an underlying fixed point. With increasing $L$, $\overline{g}$ slightly increases, consistent with decreasing $\Gamma(m=0)$ in Fig.~\ref{fig:GammaScaling_W1.5_kF0} (bottom).

For $0.3 \lesssim E \lesssim 1$, $\overline{g}$ begins to depend on $W$, and varies with $E$ by $\sim 50\%$ for $W = 1.5$ but only by $\sim 10\%$ for $W = 2.5$. For $W = 1.5$ and $E > 0.6$, $\overline{g}$ slightly decreases when $L$ grows form 100 to 200. We interpret this as a remnant of the crossover from the diffusive to the critical behavior. It is consistent that LEs obtained in this regime (not shown) cease to obey critical scaling.

{\it Discussion.}
In summary, our numerical results for DDF are consistent with localized behavior anywhere in the $m$-$E$
plane except on a critical line $m=0$, see Fig.~\ref{fig:phaseDiagram}. At $m=0$, both the dimensionless LE extrapolated to infinite system size $\Gamma_0 = 0.82 \textendash 0.85$ and the irrelevant exponent $y$ do not vary significantly with energy or disorder strength below $E\simeq 1$, while the average conductance $\overline{g}$ of fixed-size square samples at stronger disorder varies at most by $\sim 10\%$. In contrast, the localization length exponent $\nu_E$ significantly depends on energy, see Fig.~\ref{fig:phaseDiagram}(b). While $\nu_{E=0.7} = 2.53(2)$ is more or less consistent with the established value for the IQHT $\nu_\text{IQH} = 2.56 \textendash 2.62$, the value $\nu_{E}$ significantly decreases with energy down to \begin{align}
\nu_{E=0} &= \nuNull,
\label{eq:nu0_final}
\end{align} where we took a union over error bars for the two models and two scaling methods we used for $E=0$.

Let us now put our findings in the context of existing arguments and first discuss the case of large $E$ and low $W$ characterized by a large Drude conductivity $\sigma_{xx}^\mathrm{D} \gg 1$. In the SM, we numerically confirm that this regime is achievable in the DDF, albeit not for the parameters used for the scaling analysis above. Large $\sigma_{xx}^\mathrm{D}$ controls the derivation of an effective field theory for the DDF with short-range disorder \citep{Ostrovsky2007} as it justifies the required saddle point approximation. The resulting non-linear sigma model with $\theta$-term can also be derived for other models of the IQHT: the Schr\"odinger equation with short-range disorder and strong magnetic field \cite{Pruisken1984, Weidenmueller1987}, and the CC-model \cite{Zirnbauer-1997-ContinuumLimitOfCC}. These relations rationalize our finding of IQHT-like criticality in the DDF at $E = 0.7$. Note, however, that the CC-model lacks the large parameter analogous to $\sigma_{xx}^\mathrm{D}$, and the derivation of the sigma model for it is uncontrolled, as well as for the DDF at $E \simeq 0$, where $\sigma_{xx} < 1$.

We now discuss three possible scenarios addressing the $E$ dependence of $\nu_E$ [see Fig.~\ref{fig:phaseDiagram}(b)].

{\it (a) Insufficient system size.} In the history of IQHT-numerics, refined fitting functions and the ability to study larger systems shifted the value of $\nu$ considerably over time. We also cannot exclude that our results for $\nu_{E<0.7}$ are not the true asymptotic values, and further increase in $L_y$ would bring them closer to $\nu_\text{IQH}$. However, our system sizes, quality of numerical data, and its analysis are comparable to recent work on the IQHT. Also, we do not see a tendency for a drift in $\nu_{E}$ if the minimal $L_y$ involved in the fit is increased from $40$ to $68$, see SM. Finally, we corroborated our $E=0$ result \eqref{eq:nu0_final} at two disorder strengths and with an alternative scaling observable for the DDF on a lattice. Our finding for $\nu_{E=0}$ is also supported by numerical results from a massless DDF in a magnetic field \cite{Nomura2008}. At strong enough potential disorder, only the critical state deriving from the Landau level at $E=0$ persists, separating localized states at $E \lessgtr 0$. The scaling of $d\sigma_{xy}/dE|_{E=0}$ and the width of the conductance peak around $E=0$ with system size gave $\nu \approx 2.3$, but no error bars were provided.

{\it (b) Two fixed points.} In a more intriguing scenario our results could be consistent with the existence of two {\it different} fixed points. One of them is the conventional IQHT fixed point that controls the critical behavior at $E > 0$, while the other fixed point controls the system at $m=0, E=0$, see the dot in Fig.~\ref{fig:phaseDiagram}(c). We conjecture that this fixed point is multicritical, where both $m$ and $E$ are relevant, with the RG eigenvalues $y_m = 1/\nu_{E=0}$ and $y_E$. The RG flow near this point would resemble that near the tricritical point in the Ising model with vacancies \citep{Cardy-Scaling-1996}. In this scenario the critical behavior at any $E > 0$ should be the same, and coincide with that for the IQHT. Our observation of intermediate values $\nu_{E=0.3,0.5}$ may stem from the small (or even zero, if $E$ is marginally relevant) value of the crossover exponent $y_E/y_m$ at the multicritical point, resulting in the cusp-like shape of the crossover line in Fig.~\ref{fig:phaseDiagram}(c) which might cause smearing of $\nu_E$ when extracted over a too large range of $m$. However, concerns about this scenario arise from the absence of any kinks in the $\Gamma$ vs $m^2$ data for $E>0$ (see SM) as well as the apparent energy independence of $\Gamma_0$.

{\it (c) Marginal scaling.} In a recent development, Zirnbauer \cite{Zirnbauer2019} proposed a solvable conformal field theory for the IQHT, featuring a fixed point with only marginal perturbations, implying $\nu = \infty$, $y=0$. In this case, higher order terms in the $\beta$-functions for relevant and irrelevant scaling fields (the deviations $\delta\sigma_{xx}$ and $\delta\sigma_{xy}$ of the conductivities from their fixed-point values) could lead to an \emph{effective} critical exponent $\nu_\mathrm{eff}$ \cite{Zirnbauer-localisation2020-talk} dependent on the bare value of $\delta\sigma_{xx}$. For a slow RG flow of $\delta\sigma_{xx}$, $\nu_\mathrm{eff}$ could appear scale-independent but vary with the parameters of the model such as energy, see Fig.~\ref{fig:phaseDiagram}(d). Ref.~\cite{Dresselhaus2020a} reports further study of this scenario in the numerically more convenient framework of the CC-model.


{\it Outlook.} We hope our findings will prompt a careful re-examination of criticality at the IQHT and other Anderson transitions. Future work on the critical DDF should address multifractal properties of wavefunctions and compare them to established results for the IQHT \citep{EversMirlin:review}. Moreover, working with $N=3,5,7...$ flavors of DDF, the assumption $\sigma_{xx}^\mathrm{D} \gg 1$ could be justified even for $E=0$ and it would be interesting to compute $\nu_{E=0}$ in this case. Further, extension of our methods to DDF in the symmetry classes of the spin and thermal quantum Hall effects is worthwhile.


\begin{acknowledgments}
{\it Acknowledgements.} We acknowledge useful discussions with Jens Bardarson, Matt Foster, Cosma Fulga, Igor Gornyi, Alexander Mirlin, Pavel Ostrovsky, and Elio K\"onig. Computations were performed at the Ohio Supercomputer Center and the Lawrencium cluster at Lawrence Berkeley National Lab. B.S. acknowledges financial support by the German National Academy of Sciences Leopoldina through grant LPDS 2018-12 and the Quantum Science Center (QSC), a National Quantum Information Science Research Center of the U.S. Department of Energy (DOE). E.J.D was supported by NSF Graduate  Research Fellowship  Program,  NSF DGE 1752814. J.E.M. acknowledges support by TIMES at Lawrence Berkeley National Laboratory supported by the U.S. Department of Energy, Office of Basic Energy Sciences, Division of Materials Sciences and Engineering, under Contract No.\ DE-AC02-76SF00515 and a Simons Investigatorship.
\end{acknowledgments}


\setcounter{equation}{0}
\setcounter{figure}{0}

\renewcommand{\thefigure}{S\arabic{figure}}
\renewcommand{\theequation}{S\arabic{equation}}

\bibliographystyle{my-refs}
\bibliography{library-2dDiracA}

\clearpage

\appendix
\textbf{ Supplemental material for ``Criticality of two-dimensional disordered Dirac fermions in the unitary class and universality of the integer quantum Hall transition''}


\section{Non-rigorous arguments for the equivalence of IQHT and DDF criticality}

We here review the non-rigorous arguments for the original conjecture \cite{Ludwig1994} $\nu_\mathrm{E=0}=\nu_\text{IQH}$ and identify their possible flaws

(i) The original argument of Ludwig \emph{et al.}~\citep{Ludwig1994} assumes sufficiently smooth disorder in the DDF model. Then the low energy states are chiral Jackiw-Rebbi fermions moving along lines of zero mass $m + U_{z}(x,y) = 0$. Other disorders lead to random phases accumulated between saddle points in $U_{z}$, where scattering controlled by the value of $m$ occurs. This argument parallels the one that leads to the CC network model \citep{Chalker-Percolation-1988}. A possible flaw is that unlike in the large $B$-field case of the IQHT, non-chiral higher energy states might be important in DDF. Indeed, solving the Dirac equation for a linear domain wall $U_{z}(x)=cx$ arising from our smooth disorder potential with $c \sim W/a$, the first pair of counterpropagating edge modes appears at $E_{1} \sim \pm\sqrt{W}$. Naively, it takes a disorder strength $W \apprge 1$ to occupy these modes by $U_{0}$ fluctuations, but their relevance for the network model is presently unclear.

(ii) Another argument in favor for the equivalence of the CC network and DDF was given by Ho and Chalker \citep{Ho1996}. The authors showed that the spectrum of quasienergies of the clean CC model (viewed as a Floquet system) has a Dirac point with mass $m$ proportional to the deviation from the critical point. Then they argued that disorder in the CC model leads to all possible types of disorder in the Dirac model. However, the mapping assumes smooth disorder in the network model, and may be invalid for strong disorder.

\section{Quasi-1D Lyapunov exponents: Additional data, fitting procedure}

\begin{figure}[t]
\centering{}
\includegraphics{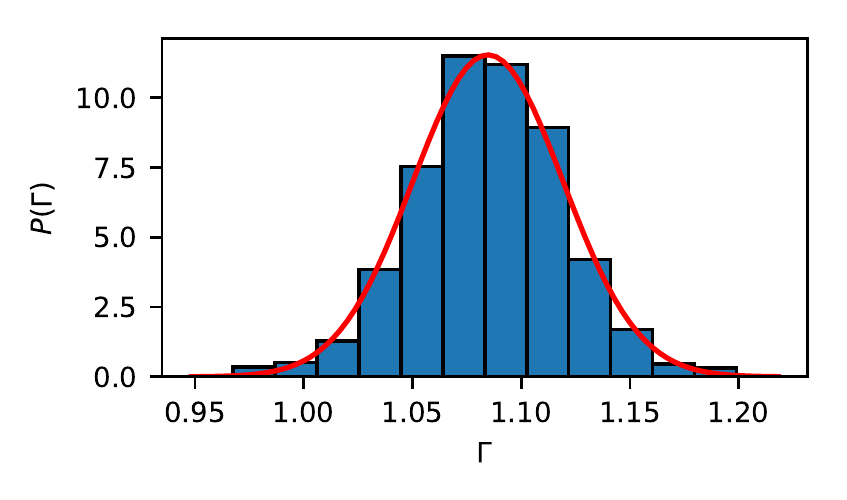}
\caption{Histogram of finite-length Lyapunov exponents for $E=0$, $W=1.5$ at $L_y=102$ and $m^2=0.05$, involving 1000 disorder realizations. The red line is a gaussian fit.
\label{fig:GammaDistribution}}
\end{figure}

\begin{figure}
\centering{}
\includegraphics{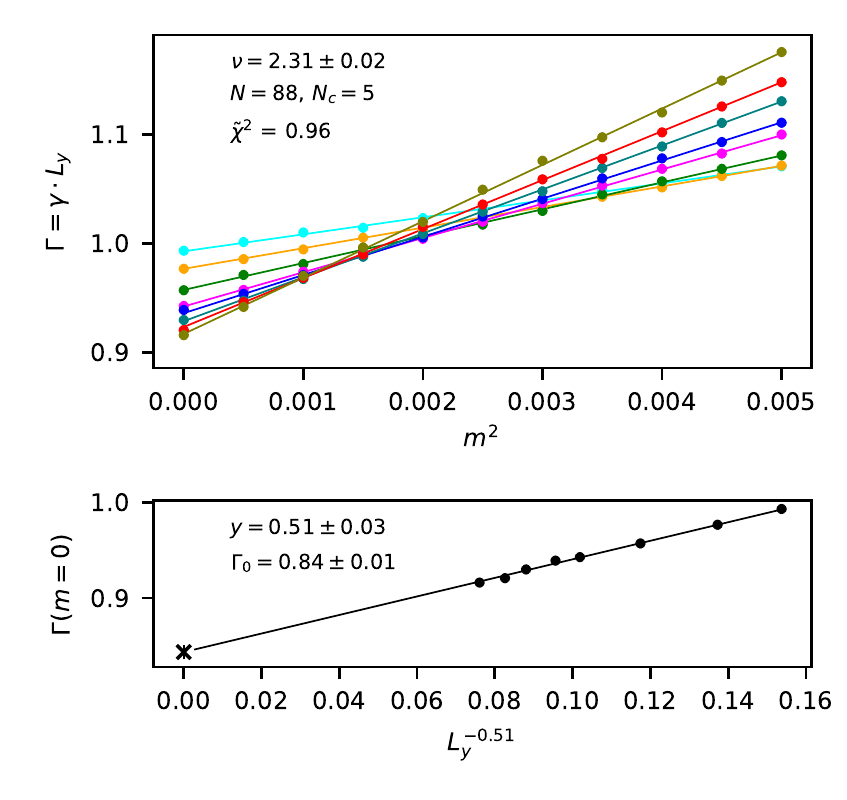}
\caption{Quasi-1D Lyapunov exponents as in Fig. 2 of the main text, but for $E=0$ and $W=2.0$. Solid lines denote the best fit with the ansatz in Eq. (3) of the main text.
\label{fig:GammaScaling_W2.0_kF0.0}}
\end{figure}

\begin{figure}
\centering{}
\includegraphics{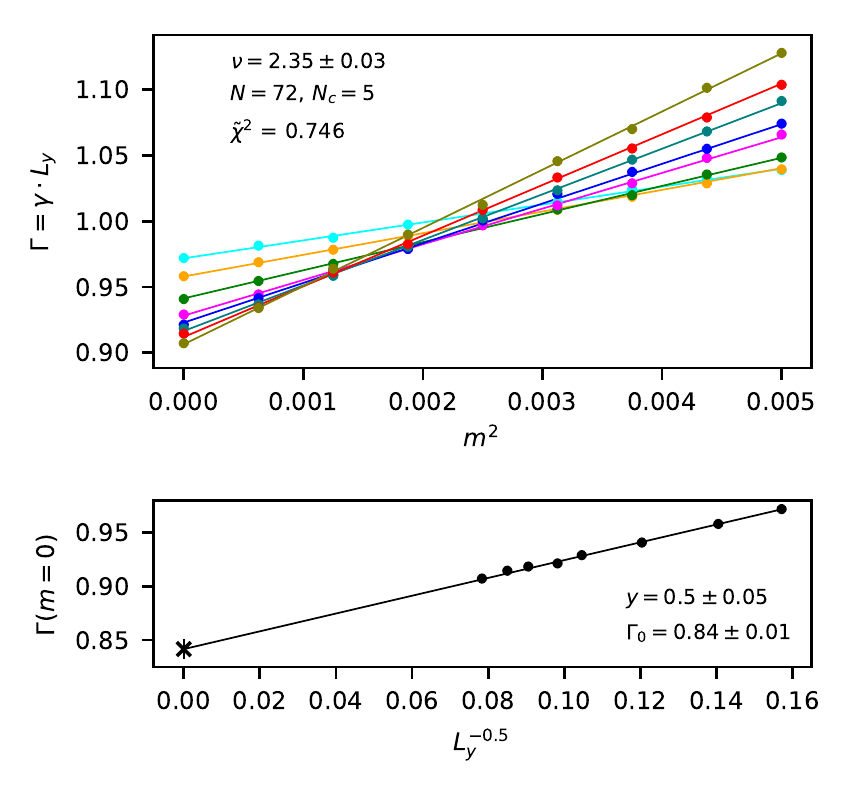}
\caption{Quasi-1D Lyapunov exponents as in Fig. 2 of the main text, but for $E=0.3$ and $W=2.0$. Solid lines denote the best fit with the ansatz in Eq. (3) of the main text.
\label{fig:GammaScaling_W2.0_kF0.3}}
\end{figure}

\begin{figure}
\centering{}
\includegraphics{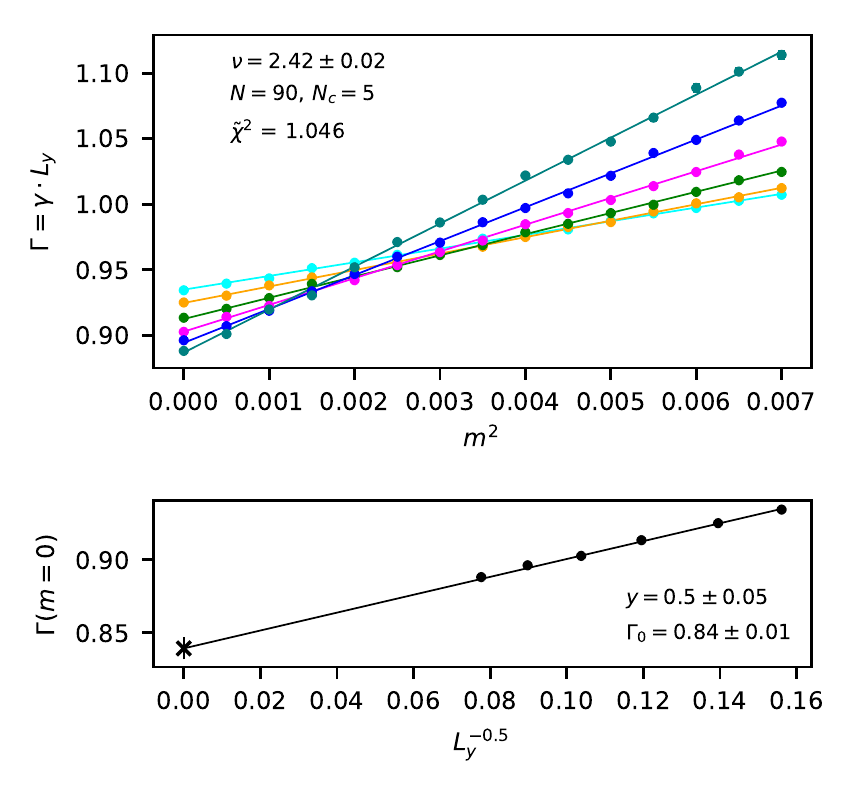}
\caption{Quasi-1D Lyapunov exponents as in Fig. 2 of the main text, but for $E=0.5$ and $W=2.0$. Solid lines denote the best fit with the ansatz in Eq. (3) of the main text.
\label{fig:GammaScaling_W2.0_kF0.5}}
\end{figure}

\begin{figure}
\centering{}
\includegraphics{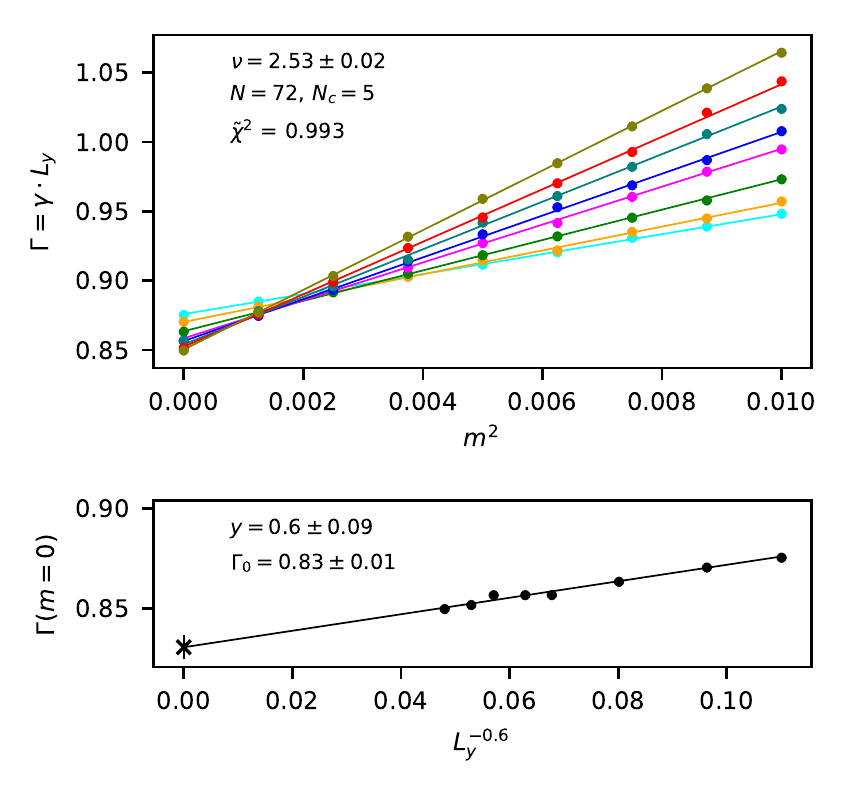}
\caption{Quasi-1D Lyapunov exponents as in Fig. 2 of the main text, but for $E=0.7$ and $W=2.0$. Solid lines denote the best fit with the ansatz in Eq. (3) of the main text.
\label{fig:GammaScaling_W2.0_kF0.7}}
\end{figure}

Here, we give a detailed description of the ensemble of data sets, fitting procedure and results for the quasi-1D Lyapunov exponents $\Gamma$ of the disordered Dirac fermion.
Table \ref{tab:LyapunovData} summarizes all data sets used in this study. The first two columns denote energy $E$ and disorder strength $W$, respectively. The third column gives the length $L_x$ of the quasi-1D slabs, the widths are $L_y=40,50,68,90,102,120,136,160$ for all data sets except for $E=0.5$, where $L_y=102,136$ are missing. We use a linearly spaced grid of squared Dirac masses $m^2$ as given in the respective column. The minimal number of disorder realizations per tuple $(m,L_y)$ is shown in the next column.
In Fig. \ref{fig:GammaDistribution} we show a typical histogram of the ensemble of finite-length $\Gamma$ approximants, for $m^2=0.05$, $L_y=102$ and $E=0$, $W=1.5$. Following Ref. \citep{Amado2011}, we confirm their gaussian distribution (red line) and approximate $\Gamma$ as the mean, the error $\sigma_\Gamma$ is the standard deviation of this ensemble divided by the square root of the number of disorder realizations. The maximal error so obtained in given in the respective column of Table \ref{tab:LyapunovData}. For a few representative data points we have increased $L_{x}$ by a factor of five and checked the $\Gamma$ only vary within error bars, moreover note the factor of three between the two values for $L_x$ used for the two data sets at $E=0$. The so obtained dimensionless Lyapunov exponents $\Gamma$ are shown as dots in Fig. 2 of the main text ($E=0$, $W=1.5$) and in Figs. \ref{fig:GammaScaling_W2.0_kF0.0}-\ref{fig:GammaScaling_W2.0_kF0.7}, for the remaining $(E,W)$ parameter pairs of table \ref{tab:LyapunovData}.

\FloatBarrier

\noindent

\begin{table*}
  \centering
\begin{tabular}{|c|c||c|c||c|c||c|c|c|c||c|c|}
\hline
{\small{}$E$} & {\small{}$W$} & {\small{}$L_{x}(\cdot10^{5})$} & {\small{}$m^{2}$} & {\small{}min realizations} & {\small{}max $\sigma_{\Gamma}/\Gamma$} & {\small{}restriction} & {\small{}$\nu$} & {\small{}$y$} & {\small{}$\Gamma_{0}$} & {\small{}$\tilde{\chi}^{2}$} & {\small{}$N$}\tabularnewline
\hline
\hline
\multirow{3}{*}{{\small{}0.0}} & \multirow{3}{*}{{\small{}1.5}} & \multirow{3}{*}{{\small{}1}} & \multirow{3}{*}{\textcolor{black}{\small{}0,0.000625,...,0.00625}} & \multirow{3}{*}{{\small{}400}} & \multirow{3}{*}{{\small{}0.2\%}} & {\small{}none} & {\small{}2.32(1)} & {\small{}0.51(3)} & {\small{}0.84(1)} & {\small{}1.060} & {\small{}88}\tabularnewline
\cline{7-12} \cline{8-12} \cline{9-12} \cline{10-12} \cline{11-12} \cline{12-12}
 &  &  &  &  &  & {\small{}$m^{2}\leq0.005$} & {\small{}2.32(1)} & {\small{}0.54(2)} & {\small{}0.85(1)} & {\small{}0.996} & {\small{}72}\tabularnewline
\cline{7-12} \cline{8-12} \cline{9-12} \cline{10-12} \cline{11-12} \cline{12-12}
 &  &  &  &  &  & {\small{}$L_{y}\geq68$} & {\small{}2.32(1)} & {\small{}0.39(9)} & {\small{}0.82(4)} & {\small{}0.978} & {\small{}66}\tabularnewline
\hline
\multirow{3}{*}{{\small{}0.0}} & \multirow{3}{*}{{\small{}2.0}} & \multirow{3}{*}{{\small{}3}} & \multirow{3}{*}{\textcolor{black}{\small{}0,0.0005,...,0.005}} & \multirow{3}{*}{{\small{}200}} & \multirow{3}{*}{{\small{}0.2\%}} & {\small{}none} & {\small{}2.31(2)} & {\small{}0.51(3)} & {\small{}0.84(1)} & {\small{}0.960} & {\small{}88}\tabularnewline
\cline{7-12} \cline{8-12} \cline{9-12} \cline{10-12} \cline{11-12} \cline{12-12}
 &  &  &  &  &  & {\small{}$m^{2}\leq0.004$} & {\small{}2.32(3)} & {\small{}0.51(5)} & {\small{}0.84(1)} & {\small{}1.163} & {\small{}72}\tabularnewline
\cline{7-12} \cline{8-12} \cline{9-12} \cline{10-12} \cline{11-12} \cline{12-12}
 &  &  &  &  &  & {\small{}$L_{y}\geq68$} & {\small{}2.30(3)} & {\small{}0.47(10)} & {\small{}0.84(2)} & {\small{}0.919} & {\small{}66}\tabularnewline
\hline
\hline
\multirow{3}{*}{{\small{}0.3}} & \multirow{3}{*}{{\small{}2.0}} & \multirow{3}{*}{{\small{}1}} & \multirow{3}{*}{{\small{}0,0.000625,...,0.005}} & \multirow{3}{*}{{\small{}400}} & \multirow{3}{*}{{\small{}0.2\%}} & {\small{}none} & {\small{}2.35(3)} & {\small{}0.50(5)} & {\small{}0.84(1)} & {\small{}0.746} & {\small{}72}\tabularnewline
\cline{7-12} \cline{8-12} \cline{9-12} \cline{10-12} \cline{11-12} \cline{12-12}
 &  &  &  &  &  & {\small{}$m^{2}\leq0.004375$} & {\small{}2.37(3)} & {\small{}0.53(5)} & {\small{}0.85(1)} & {\small{}0.756} & {\small{}64}\tabularnewline
\cline{7-12} \cline{8-12} \cline{9-12} \cline{10-12} \cline{11-12} \cline{12-12}
 &  &  &  &  &  & {\small{}$L_{y}\geq50$} & {\small{}2.35(3)} & {\small{}0.44(9)} & {\small{}0.83(2)} & {\small{}0.721} & {\small{}63}\tabularnewline
\hline
\multirow{3}{*}{{\small{}0.5}} & \multirow{3}{*}{{\small{}2.0}} & \multirow{3}{*}{{\small{}2}} & \multirow{3}{*}{\textcolor{black}{\small{}0,0.0005,...,0.007}} & \multirow{3}{*}{{\small{}100}} & \multirow{3}{*}{{\small{}0.3\%}} & {\small{}none} & {\small{}2.42(2)} & {\small{}0.50(5)} & {\small{}0.84(1)} & {\small{}1.046} & {\small{}90}\tabularnewline
\cline{7-12} \cline{8-12} \cline{9-12} \cline{10-12} \cline{11-12} \cline{12-12}
 &  &  &  &  &  & {\small{}$m^{2}\leq0.006$} & {\small{}2.44(3)} & {\small{}0.53(5)} & {\small{}0.84(1)} & {\small{}1.049} & {\small{}78}\tabularnewline
\cline{7-12} \cline{8-12} \cline{9-12} \cline{10-12} \cline{11-12} \cline{12-12}
 &  &  &  &  &  & {\small{}$L_{y}\geq68$} & {\small{}2.40(4)} & {\small{}0.59(18)} & {\small{}0.85(2)} & {\small{}1.051} & {\small{}60}\tabularnewline
\hline
\multirow{2}{*}{{\small{}0.7}} & \multirow{2}{*}{{\small{}2.0}} & \multirow{2}{*}{{\small{}1}} & \multirow{2}{*}{{\small{}0,0.00125,...,0.01}} & \multirow{2}{*}{{\small{}550}} & \multirow{2}{*}{{\small{}0.2\%}} & {\small{}none} & {\small{}2.53(2)} & {\small{}0.60(9)} & {\small{}0.83(1)} & {\small{}0.993} & {\small{}72}\tabularnewline
\cline{7-12} \cline{8-12} \cline{9-12} \cline{10-12} \cline{11-12} \cline{12-12}
 &  &  &  &  &  & {\small{}$m^{2}\leq0.0075$} & {\small{}2.55(4)} & {\small{}0.57(10)} & {\small{}0.83(1)} & {\small{}1.001} & {\small{}56}\tabularnewline
\hline
\end{tabular}{\small\par}
  \caption{Overview of quasi-1D Lyapunov exponent data sets and fitting results for the best fitting function, Eq. (3) of the main text. The set of system widths is $L_y=40,50,68,90,102,120,136,160$ for all data sets except for $E=0.5$, where $L_y=102,136$ are missing. The range of system sizes has been restricted for a stability check as indicated.}
  \label{tab:LyapunovData}
\end{table*}

\begin{figure}
\centering{}
\includegraphics{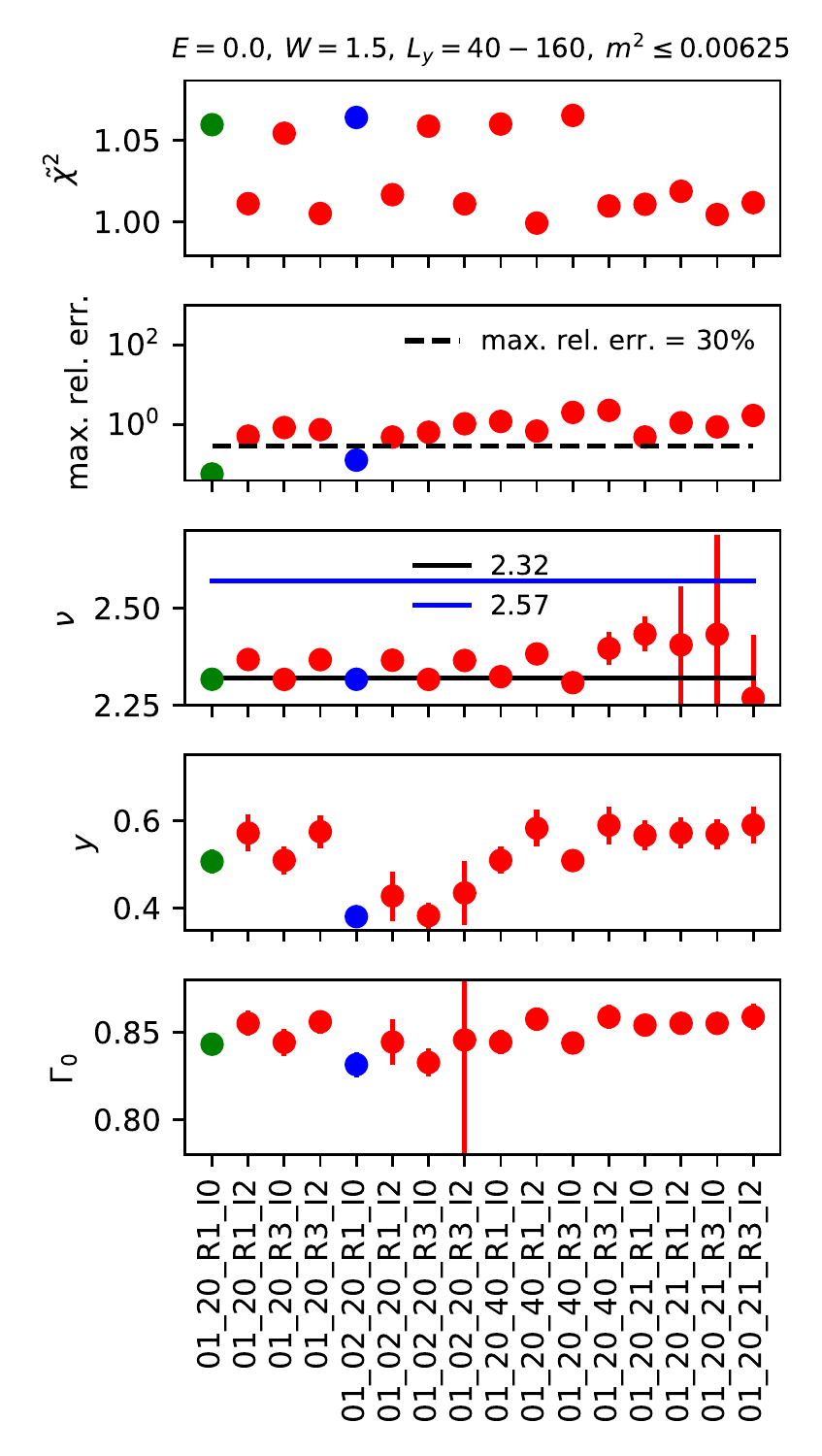}
\caption{Stability plot showing the performance of different fitting functions for the quasi-1D Lyapunov exponents obtained for $E=0$ and $W=1.5$ with all available $(m,L_y)$ data points. The labels on the $x$-axis encode the chosen fitting function, c.f. Eq. \eqref{eq:GammaSeries}. The green color indicates the best fit for which the maximum relative error of all fit parameters is below $30\%$, see dahsed line in the second panel. In the third panel, horizontal lines denote $\nu=2.57$ as obtained recently for the CC network model in Ref. \cite{Klumper2019} and our estimate $\nu_{E=0} = 2.32$.
\label{fig:stabilityPlot_E0_W1.5}}
\end{figure}

\begin{figure}
\centering{}
\includegraphics{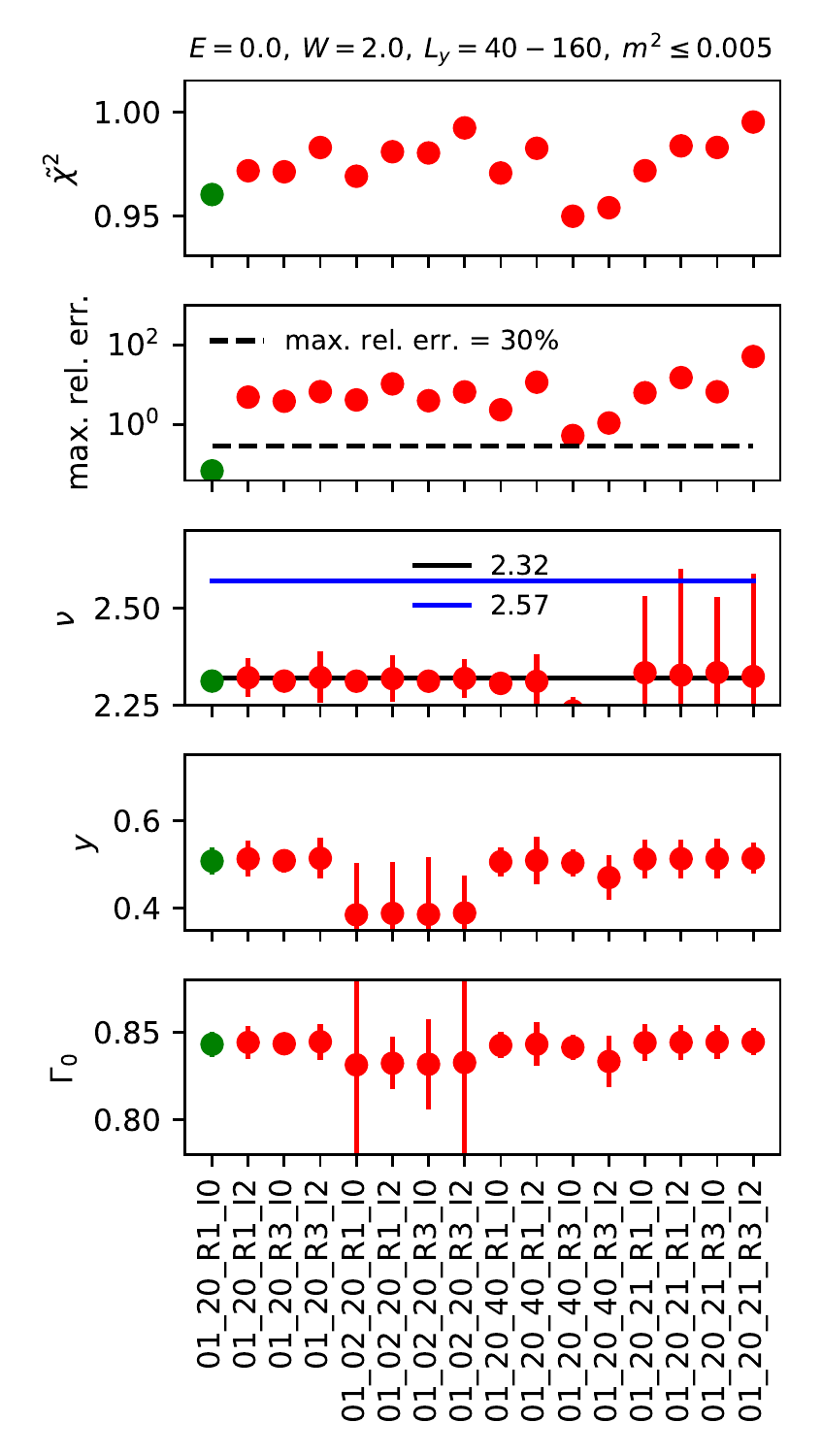}
\caption{Stability plot as in Fig. \ref{fig:stabilityPlot_E0_W1.5}, but for $E=0$, $W=2$.
\label{fig:stabilityPlot_E0_W2}}
\end{figure}

\begin{figure}
\centering{}
\includegraphics{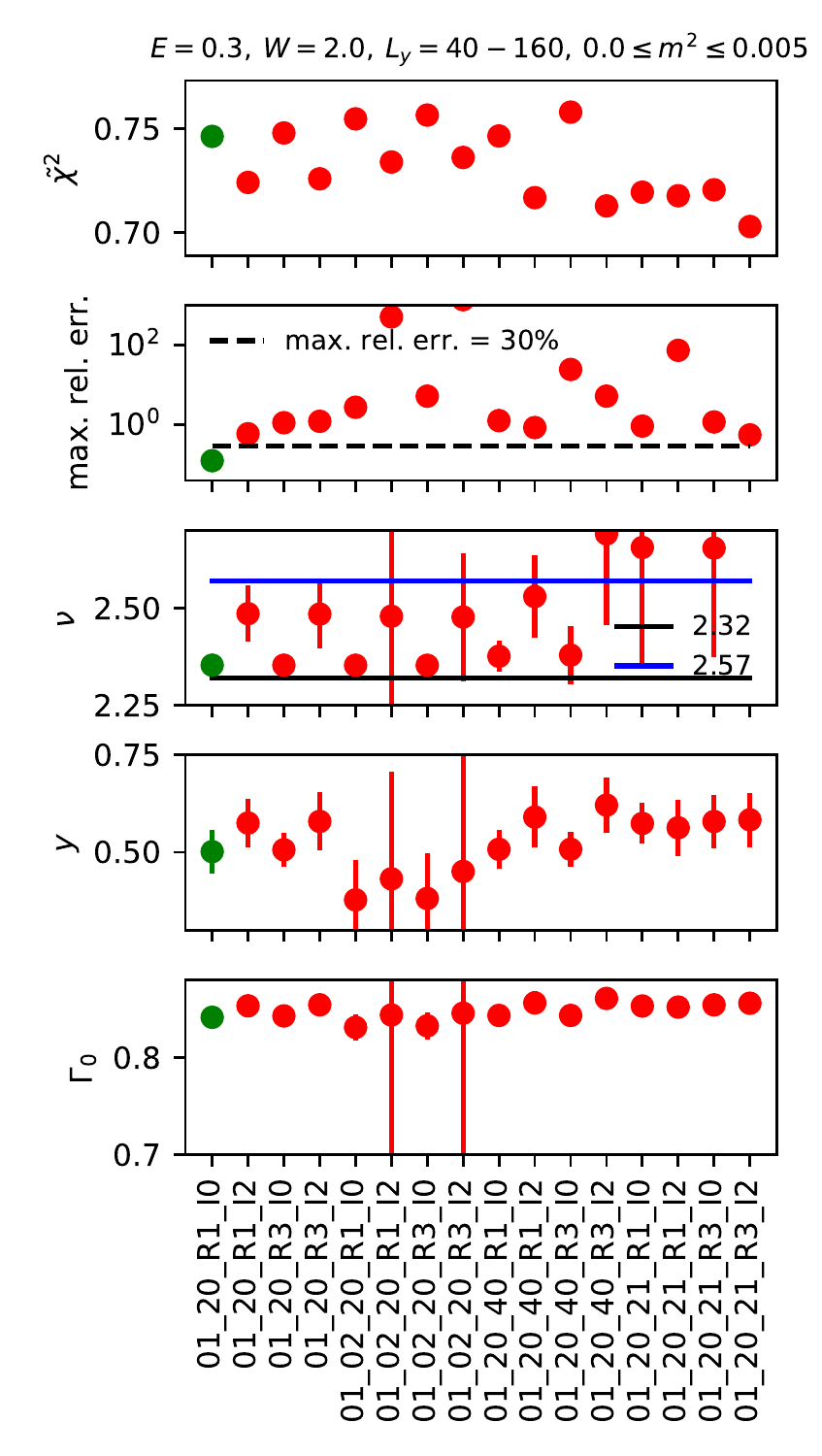}
\caption{Stability plot as in Fig. \ref{fig:stabilityPlot_E0_W1.5}, but for $E=0.3$, $W=2$.
\label{fig:stabilityPlot_E0.3_W2}}
\end{figure}

\begin{figure}
\centering{}
\includegraphics{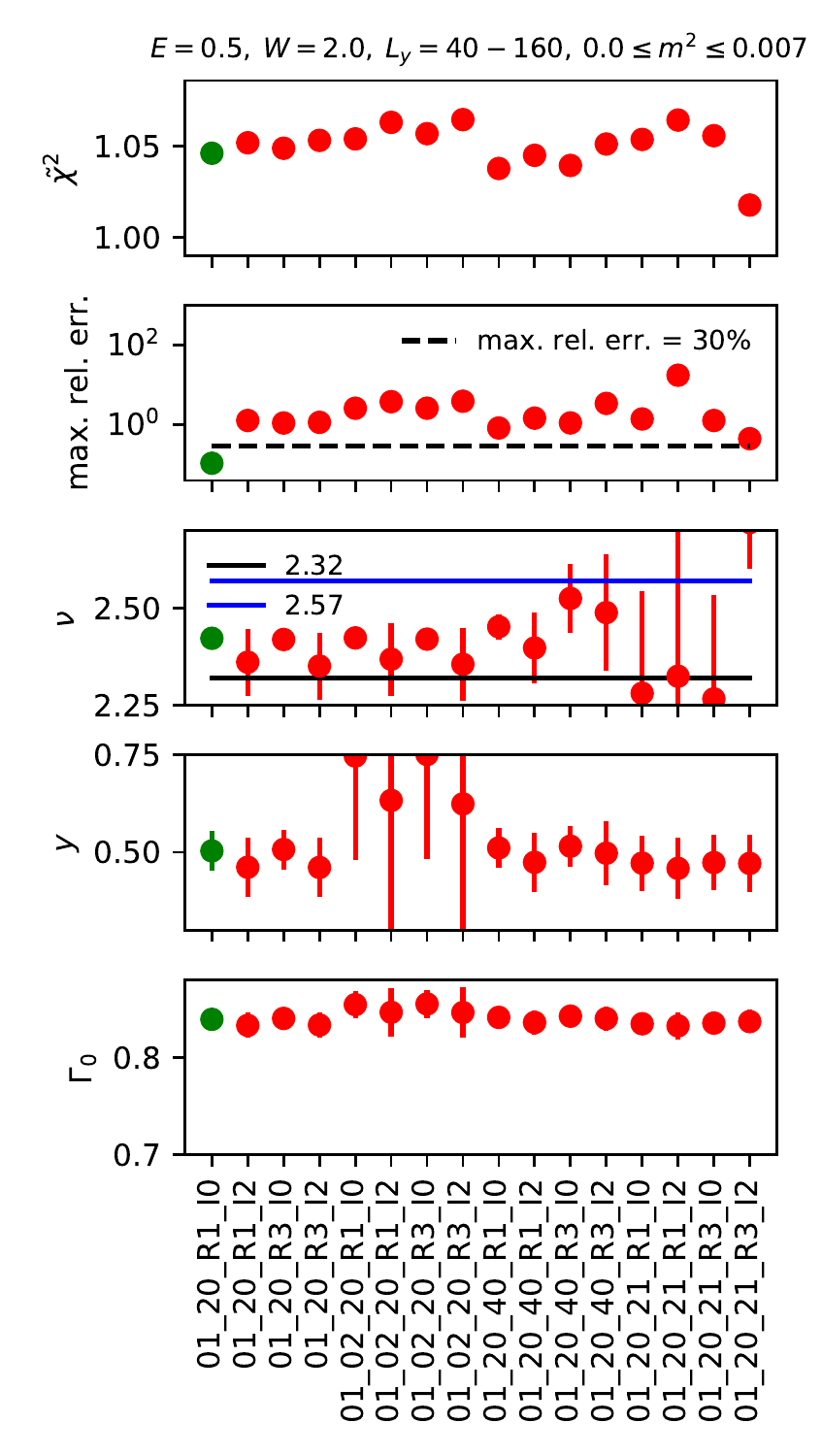}
\caption{Stability plot as in Fig. \ref{fig:stabilityPlot_E0_W1.5}, but for $E=0.5$, $W=2$.
\label{fig:stabilityPlot_E0.5_W2}}
\end{figure}

\begin{figure}
\centering{}
\includegraphics{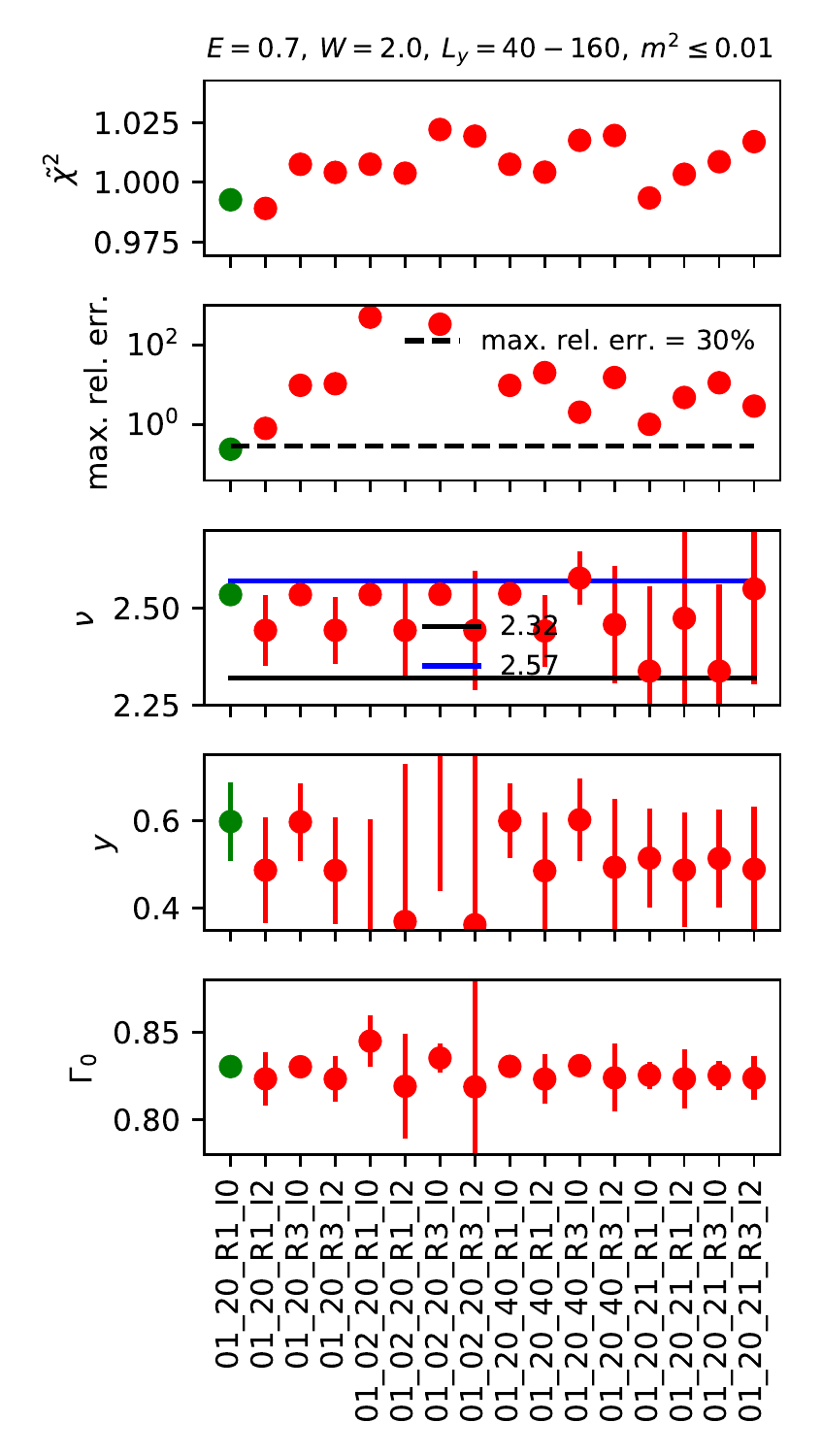}
\caption{Stability plot as in Fig. \ref{fig:stabilityPlot_E0_W1.5}, but for $E=0.7$, $W=2$.
\label{fig:stabilityPlot_E0.7_W2}}
\end{figure}

We now consider the scaling ansatz for fitting the $\Gamma(m,L_y)$ data. Following Ref. \citep{Slevin-Finite-2012}, we take into account the relevant and the leading irrelevant scaling variables,
$x_{R}(m)$ and $x_{I}(m)$, respectively:
\begin{align}
\Gamma(m,L_y) & = F \left( X_{R}=x_{R}(m)L_y^{1/\nu},X_{I}=x_{I}(m)L_y^{-y} \right).
\label{eq:GammaAnsatz}
\end{align}
The relevant variable vanishes at the critical point, $x_{R}(0)=0$. To satisfy the (statistical) symmetry $m \rightarrow -m$, we follow the customary choice \citep{Slevin-Finite-2012, Amado2011} an expand $x_{R}(m)$ as an odd-power polynomial of order $R$, and $x_{I}(m)$ as an even-power polynomial of order $I$. Then the series expansion of Eq. \eqref{eq:GammaAnsatz} must contain only even powers of $X_{R}$:
\begin{equation}
\Gamma(m,L_y) = \Gamma_{0} + \alpha_{01} X_{I} + \alpha_{20} X_{R}^{2} + \alpha_{02} X_{I}^{2} + \alpha_{21} X_{R}^{2} X_{I} + ...
\label{eq:GammaSeries}
\end{equation}
We fix the expansion orders $R,I$ and the truncation of the series
in Eq. \eqref{eq:GammaSeries} and find the parameters $\nu,y,\Gamma_{0}, \alpha_{ij}$ using non-linear least squares fitting. The quality of a fit is reported by the value of $\tilde{\chi}^{2}=\chi^{2}/(N-N_{c})$ where $N$ is the number of data points $\Gamma_j$ (j stands for the tuple $(m,L_y)$), $N_{c}$ is the number of fitting parameters in the chosen scaling function, and $\chi^{2} = \sum_{j=1}^{N} (\Gamma_{j}-\Gamma)^{2}/\sigma_{j}^{2}$ is the sum of squared residues weighted by the variances $\sigma_{j}^{2}$. The errors of the fit parameters (one standard deviation) are obtained from repeated fitting of $\sim100$ synthetic data sets generated from a normal distribution of mean $\Gamma_{j}$ and variance $\sigma_{j}^{2}$.
The fit results, i.e. the optimized universal parameters of the ansatz with the overall lowest $\tilde{\chi}^{2}$ and all relative errors of fit parameters below $30\%$, are given in the last three columns of Table \ref{tab:LyapunovData} and also in Fig. 2 and  \ref{fig:GammaScaling_W2.0_kF0.0}-\ref{fig:GammaScaling_W2.0_kF0.7}. It turns out that for all data sets considered, the minimal ansatz which allows for an irrelevant contribution (first three terms on the rhs of Eq. \ref{eq:GammaSeries} with $R=1$, $I=0$ and $N_c=5$, given explicitly in Eq.~(3) of the main text) is chosen by our fitting algorithm. As shown in the stability plots of Figs. \ref{fig:stabilityPlot_E0_W1.5} to \ref{fig:stabilityPlot_E0.7_W2}, while sometimes a higher order fitting function can give a slightly smaller value of $\tilde{\chi}^2$, this always comes at the cost of a relative error exceeding our stabilitiy bound above. As a further stability criterion, we request that with one or two small system widths $L_y$ or a few largest masses $m$ removed from the fit, critical properties still overlap in error bars. This is also confirmed in Table \ref{tab:LyapunovData}, see column titled ``restriction''.

\noindent

\noindent

\noindent

\noindent

\noindent

\noindent

\noindent

\noindent

\noindent

\clearpage
\section{Details for calculation and fitting of the alternative scaling observable}

Here, we provide details for the calculation of the alternative scaling observable $\Lambda$ proposed by Fulga \textit{et al.} \citep{Fulga2011a}. For tight-binding models, the generalized twisted transversal boundary conditions mentioned in the main text can be applied using the concept of virtual leads. While we refer the reader to Ref. \citep{Fulga2011a} for details, we remark that caution has to be used when defining virtual leads for a tight-binding model with multiple orbitals per site. In order to apply the generalized twisted BC $\psi_{x,y=L-1} = z\, \psi_{x,y=0}$ as given in the main text on the level of the scattering matrix, one has to ensure that the scattering matrix is defined with respect to the same propagating modes in both virtual leads at the top and the bottom. This is not automatically fulfilled in \noun{kwant} so that an appropriate unitary rotation is required to post-process the computed scattering matrix before applying the twisted BC.

\begin{figure}[t]
\centering{}
\includegraphics{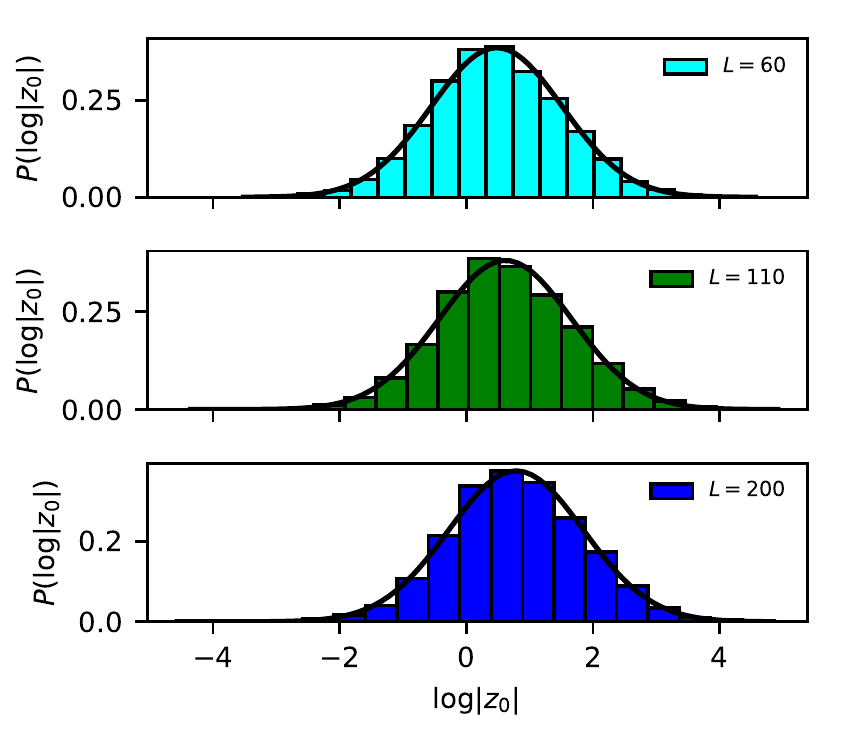}
\caption{{\small{}\label{fig:histLogz0}}Histograms for the quantity $\mathrm{log}|z_0|$ underlying Fig. 3 of the main text for $m=0.05$. The solid lines are Gaussians for comparison.}
\end{figure}

As explained in the main text, we find the $z=z_0$ where the determinant of the reflection matrix $r$ vanishes. In Fig. \ref{fig:histLogz0} we show the histogram of $\mathrm{log}|z_0|$ for subcritical $m=0.05$ and three different system sizes. Like in the case of Lyapunov exponents, they are gaussian distributed and their mean and standard deviation determine the scaling observable $\Lambda$ reported in Fig. 3 of the main text.

The fitting procedure and scaling function $\Lambda(m,L)$ parallels the discussion for the Lyapunov exponents $\Gamma(m,L_y)$ discussed above. However, as there is no symmetry fixing $m_c=0$, we have to introduce a fitting parameter $m_c$ so that $\delta=m-m_{c}$ appears in the expansion of the relevant scaling field and general polynomials are allowed. We find the ansatz $\Lambda(m,L)=\beta_{1}X_{R}+\beta_{3}X_{R}^{3}$ with $X_{R}=L^{1/\nu}[\delta+\alpha\delta^{2}])$ to be optimal in the sense defined above. The fit in Fig. 3 of the main text is characterized by $N_c=5$, $N=50$ and $\tilde{\chi}^2=0.977$. In particular, this reasonable fitting result does not require the inclusion of an irrelevant contribution, although we have not found a symmetry reason for this. For the CC model, the same was observed in Ref. \citep{Fulga2011a}. However, we want to emphasize that smaller system sizes or other models might well require an irrelevant contribution in the scaling function.

We finally remark that Fulga \textit{et al.} \citep{Fulga2011a} claim that their scaling observable requires samples of large aspect ratio. However, we find that calculations with square samples yield converging numerics and save considerable computational resources. We discern no underlying argument in favor of particular aspect ratios.

\noindent

\section{Landauer conductance for square system at $m=0,E=0$}

Critical conductance distributions depend on the sample shape and boundary conditions, and are typically very broad \citep{Kramer-Random-2005, EversMirlin:review}. In addition, unlike the self-averaging Lyapunov exponents, critical transport observables exhibit multifractality: their different moments are described by different scaling dimensions \citep{Janssen-1999, Klesse-2001, Gruzberg-Classification-2013}.

\begin{figure}
\centering{}
\includegraphics{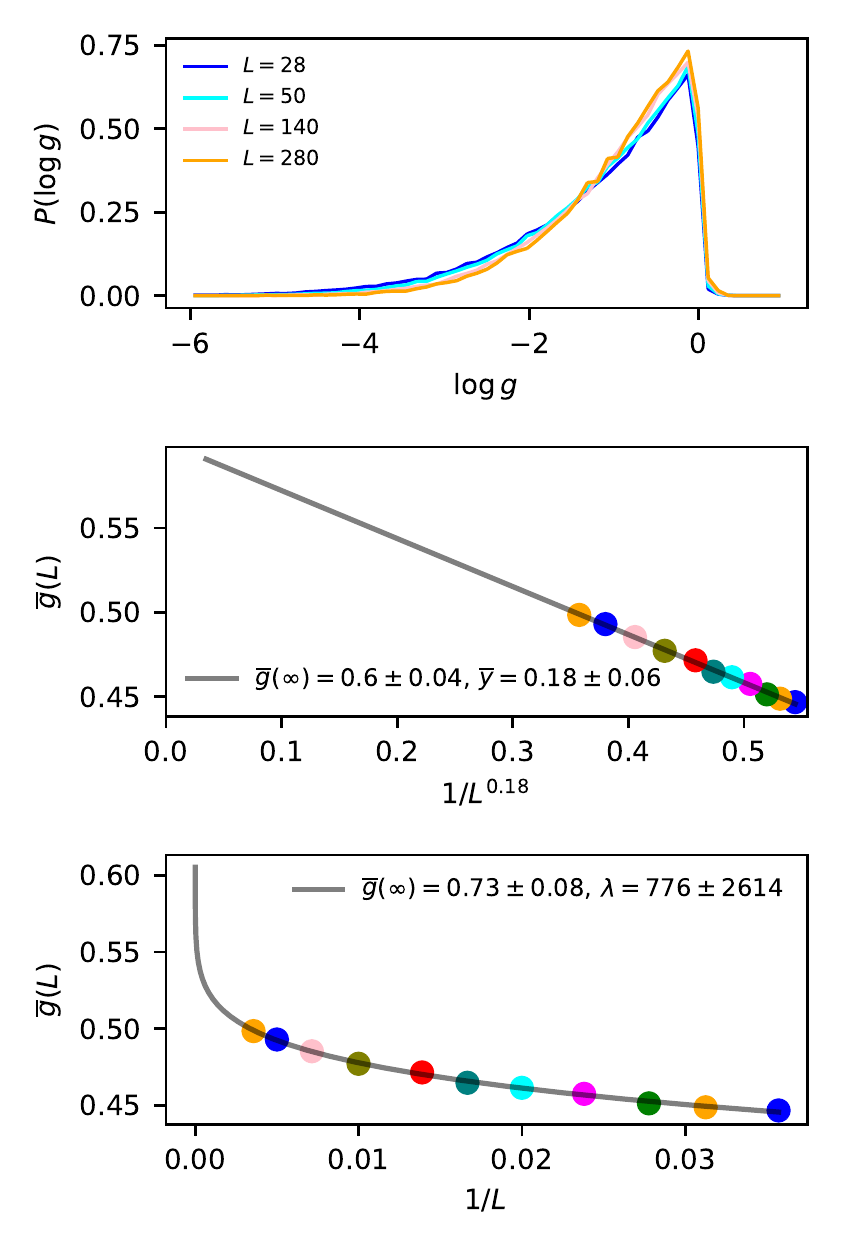}
\caption{{\small{} Landauer conductance of square continuum Dirac systems, Eq. (3) of the main text, with $m=0$, $E=0$ and smooth disorder, Eq. (4), of strength $W=2.5$. The top panel shows the histogram of $\mathrm{ln}g$ for four system sizes between $L=28$ and $280$ based on between $80000$ and $20000$ disorder realizations. The middle panel shows the mean square conductance (with additional intermediate system sizes) with a fit (solid line) to the power-law $\overline{g}(L)=\overline{g}(\infty) - a L^{-\overline{y}}$ with fit parameters given in the panel. The bottom panel shows the same data with a logarithmic fit
$\overline{g}(L)=\overline{g}(\infty) - b\, \mathrm{log}(\lambda L)$.
The solid line indicates a power law fit as described in the main text.}}
\label{fig:gsq}
\end{figure}

In the main text, we have presented disorder averaged square system Landauer conductance $\overline{g}$ for $m=0$ and a range of energies. Here, focusing on the case $E=0$, we present the histogram of $\mathrm{log}g$, see Fig. \ref{fig:gsq} (top) for a range of system sizes $L = 28 \textendash 280$. It is qualitatively similar to the IQHT case \citep{Jovanovic1998,Schweitzer2005}.

In Fig. \ref{fig:gsq} (middle), we attempt a power law fit of the average conductance, using the ansatz $\overline{g}(L) = \overline{g} - \alpha L^{-\bar{y}}$. It yields $\bar{y} = 0.18(6)$ and $\overline{g} = 0.60(4)$. We emphasize that despite the large number of disorder realizations, the fit is very instable and we were not able to find meaningful error bars for the data points given the broad and unknown probability distribution of $g$. The bottom panel presents a logarithmic fit $\overline{g}(L)=\overline{g}(\infty) - b\, \mathrm{log}(\lambda L)$ as has been suggested in literature as well \citep{Puschmann-Integer-2019}.

\noindent

Notice that due to the above-mentioned multifractality of conductances at criticality, we do not expect that the irrelevant exponent $\overline{y}$ describing the approach of $\overline{g}(L)$ to its limiting value is the same as the exponent $y$ for the self-averaging Lyapunov exponent $\Gamma$, which is related to the conductance of a strongly localized quasi-1D system.


\noindent
\begin{figure}[t]
\centering{}
\vspace{2mm}
\includegraphics{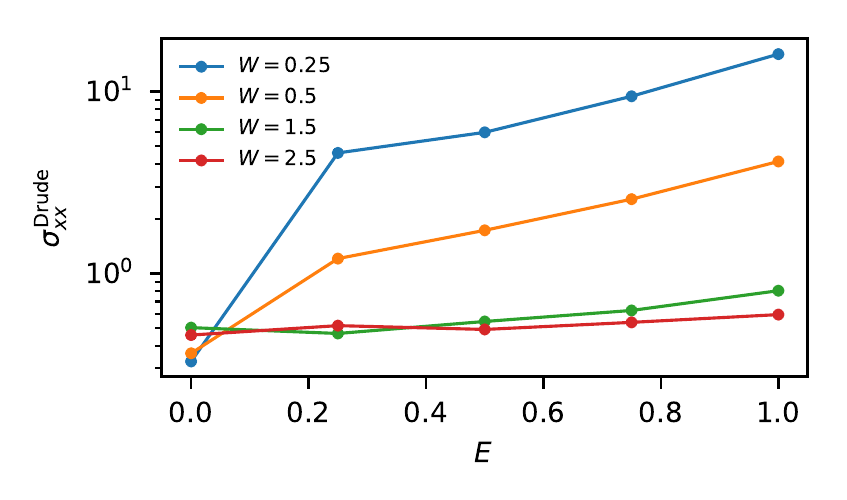}
\caption{\small{} Drude conductivity $\sigma_{xx}^\mathrm{Drude}$ as a function of energy $E$ and disorder strength $W$ for the DDF continuum model with smooth disorder. The data is obtained from conductance simulations for samples of large aspect ratio $L_y \gg L_x$ as the inverse slope of resistance curves $\overline{r}(L_x)=1/\overline{g}(L_x)$ for $L_x$ in the range of a few mean free paths. The mean free path is estimated by comparison to the conductance data of the clean ballistic case $W=0$.}
\label{fig:Drude}
\end{figure}

\section{Drude conductivity}

In Ref. \citep{Ostrovsky2007}, Ostrovsky \textit{et al.} derived Pruisken's non-linear sigma model as an effective long-range theory for the DDF. As emphasized in the main text, this derviation is only controlled if the longitudinal Drude conductivity $\sigma_{xx}^\mathrm{D}$ is much larger than unity. In Fig. \ref{fig:Drude}, we plot the numerically obtained Drude conductivity for the continuum DDF with smooth disorder at $m=0$ for a range of energies $E$ and disorder strengths $W$, see caption for details of the procedure. In the metallic limit of large $E$ and small $W$, the data obeys $\sigma_{xx}^\mathrm{Drude} \sim 1/W^2$ \citep{Ostrovsky2007}. For the parameter combinations $(E,W)$ used for the scaling analysis above, the Drude conductivity in Fig. \ref{fig:Drude} is of order unity such that the derivation of Ref. \citep{Ostrovsky2007} is not controlled.

\end{document}